 \documentclass[aps,prb,twocolumn,superscriptaddress,floatfix]{revtex4}

\usepackage{graphicx}
\usepackage{bm}

\begin{document}


 \title{Does confining the hard-sphere fluid
   between hard walls change its average properties?}


 \author{Jeetain Mittal}
\email[]{jeetain@che.utexas.edu}
 \affiliation{Department of Chemical Engineering, The University of Texas at 
 Austin, Austin, TX}

 \author{Jeffrey R. Errington}
\email[]{jerring@buffalo.edu}
 \affiliation{Department of Chemical and Biological Engineering, University at 
 Buffalo, The State University of New York, Buffalo, NY}

 \author{Thomas M. Truskett}
 \email[]{truskett@che.utexas.edu}
 \affiliation{Department of Chemical Engineering, The University of Texas at 
 Austin, Austin, TX}
 \affiliation{Institute for Theoretical Chemistry, The University of Texas at Austin, Austin, TX}


 \date{\today}

\begin{abstract}
We use grand 
canonical transition-matrix Monte Carlo and discontinuous molecular
dynamics simulations to generate precise
thermodynamic and kinetic data for 
the equilibrium hard-sphere fluid confined between
smooth hard walls.  These simulations show that the pronounced inhomogeneous
structuring of the fluid normal to the confining walls, 
often the primary focus of density functional theory studies,
has a negligible effect on many of its average 
properties over a
surprisingly broad range of conditions. 
We present one consequence of this
insensitivity to confinement:  a simple analytical equation relating 
the average density of the confined fluid to that of the bulk fluid with 
equal activity.  Nontrivial implications 
of confinement for 
average fluid properties do emerge in this system, but only when
the fluid is {\em both} (i) dense {\em and} (ii) confined to a 
gap smaller than approximately three particle diameters.  For this limited 
set of conditions, we find that 
``in-phase'' oscillatory deviations in excess
entropy and self-diffusivity (relative to the behavior of the bulk fluid 
at the same average density) occur as a function of gap size.  
These paired thermodynamic/kinetic deviations from bulk behavior 
appear to reflect the geometric packing frustration that arises when the confined space 
cannot naturally accommodate an integer number of particle layers.  
\end{abstract}
\maketitle
%
%
\section{Introduction \label{Intro}} 
 Confined fluids play an important role in a host of scientific phenomena and 
technological applications. 
Examples range from the aqueous fluids that fill the cytostructures of
biological cells to the solvents that facilitate the operation of
nano- and microfluidic devices, membranes for separations, 
and porous catalytic materials. 
In many of these systems, confinement significantly modifies the 
thermodynamic and kinetic behavior of the
fluid relative to the bulk phase. 
Such modifications 
are generally attributed to the collective effects of the
size and shape of the confined space and the interactions of the fluid 
with the confining surfaces. 
However, isolating 
the individual contributions of these various factors for study 
can be a daunting
experimental task.

Given this difficulty, one alternative approach 
has been to explore the behavior of simplified models that allow
one to examine the implications of confinement in the absence
of the complicating details that are present in experimental systems.  
Along these lines, a commonly 
investigated model is the equilibrium, monatomic hard-sphere (HS) fluid 
confined between smooth and parallel hard walls. 
This is arguably the most basic model that can capture
the main entropic packing effects associated with fluids in
confined spaces.  Its characteristic inhomogeneous density 
profile (normal to the confining walls), which has been 
a primary focus of previous investigations, is now qualitatively understood.\cite{htdavis}~
Unfortunately, despite progress in elucidating some of the 
other properties of this system,
\cite{htdavis,marjolein2006,robbins1992,matthias1996,rice1998,lowen2dp,mittal2006,marjolein2004,auer2003,kegel2001,heni1999}~
a comprehensive picture for precisely how confinement modifies 
the average thermodynamic and kinetic behavior of the equilibrium 
HS fluid has yet to emerge.

One of the most basic hurdles to constructing this picture 
has been the lack of accurate molecular simulation data for the average
properties of the confined
HS fluid, a fact that may seem surprising given the apparent simplicity of 
the model.  Ironically, the model's simplicity has 
indirectly contributed to the lack of simulation data because it has
allowed the system to be readily studied by approximate 
theories instead,\cite{ted1985,ted1987}~which are appealing 
because they are physically insightful and require only 
modest computational resources.  However, the development of efficient 
algorithms 
for investigating systems with discontinuous potentials
and the availability of fast computers have now made it
feasible to use molecular simulation to fully characterize the behavior of 
this model 
with both accuracy and precision.  One aim 
of the present study is to leverage these simulation 
resources to take an important
step toward completing this characterization.

The data that we present here provide some insights
into an important, but still poorly understood, conceptual point concerning this model.
Specifically, it has not been entirely clear how one should 
compare the confined fluid to the 
bulk fluid in order to elucidate the main effects of confinement.
One obvious possibility is to compare the two systems under conditions where they
exhibit equal ``average'' density.  The argument for choosing
this basis of comparison 
is straightforward.
Packing effects dominate the behavior of athermal systems, and average 
density is an important factor in determining how the particles pack.  
Moreover, if one controls for average density in making the
comparison, then one can hope to isolate more
subtle effects due to, e.g., the finite
size of the confined system (in one direction) and the shape of the density
profile (i.e., the ``layering'').
Another possibility is to compare the two systems at equal activity, where 
the bulk and pore fluids exhibit different average densities.  The 
advantage in doing so is also obvious.  Equality of activity is a relevant 
experimental constraint on the chemical equilibrium that is established 
between the bulk and pore fluids.

The complication in comparing the two systems at the same density is that 
there are two different definitions for average density that
are commonly invoked: 
$\rho=N\sigma^3/V$ and $\rho_h=N\sigma^3/V_h$. Here, 
$N$ refers to the number of particles, and $\sigma$ is the particle
diameter.  The difference between the two is that 
$V=AH$ is the total volume of the confined fluid 
(i.e., $A$ is the area of a wall in contact with the fluid, and $H$ is
the distance from ``wall surface to wall surface''), while $V_h=Ah$ is
the smaller volume 
accessible to the particle centers (i.e., $h=H-\sigma$).  While
densities based on these two definitions converge in the limit $H
\rightarrow \infty$, they can be quite
different for severely confined fluids.  
We are not aware of any 
systematic comparisons for how the thermodynamic properties of this
system depend on $\rho$ and $\rho_h$, respectively.

However, 
even in the absence of such comparitive studies, it is easy
to imagine that one might indeed arrive at
qualitatively different conclusions about the implications of confinement
depending on whether $\rho$ or $\rho_h$ is chosen as the basis
for comparison.  To appreciate this point, 
consider that $\rho_h$ diverges in the limit where 
the gap size $H$ is reduced, at fixed $N/A$, 
to the size of one particle 
diameter $\sigma$ (i.e., the two-dimensional fluid limit), whereas $\rho$
and many other fluid properties of interest remain finite.  
This type of consideration alone hints that $\rho$ might be the more
suitable density variable of the two for making comparisons to the
bulk fluid, and indeed $\rho$ naturally 
emerges in the thermodynamic analysis of confined 
HS fluids.\cite{rice1998,lowen2dp}~

More concrete evidence supporting the use of 
$\rho$ rather than $\rho_h$ for comparing confined and bulk fluids comes from studies of transport
properties.  Specifically, it has recently been 
demonstrated via molecular simulation\cite{mittal2006}~
that the self-diffusivity of the
confined HS fluid parallel to the confining walls, over a broad range
of equilibrium conditions, is very similar to the diffusion coefficient 
of the bulk HS fluid if the two systems are compared at the same 
value of $\rho$.  In other words, the specific details of the 
inhomogeneous packing structures have only minor influence on the average 
single-particle dynamics of the confined fluid, as long
as one controls for the average overall density $\rho$.  
Alternatively, if one instead compares the behaviors of the 
bulk and confined systems at
equal values of $\rho_h$, one arrives at the 
conclusion that confining a HS fluid between hard 
walls has the effect of significantly 
speeding up its dynamics.
This latter artificial conclusion is related to the fact that $N/A$
must vanish if $\rho_h$ is to remain constant in the limit $H
\rightarrow \sigma$. As a result, even if the numerical value of
$\rho_h$ is chosen to be indicative of a dense bulk fluid, 
the actual average interparticle separation and particle 
mobility in the lateral direction 
will generally be very large (e.g., comparable to a dilute gas) when
the fluid is confined to small enough $H$.

In this paper, we follow up on some of 
these initial observations by presenting
a more comprehensive study for how fluid density and confinement
(between hard walls)
affect the thermodynamic and kinetic
properties of the HS fluid.  We broadly focus our investigation 
on four main questions.  The first
pertains to the 
equation of state of the confined fluid (i.e., how the 
average transverse and normal components of its pressure
tensor vary with average density).  Specifically, we are interested 
in how the behaviors of these pressure components 
depend on the volume definition invoked, i.e., $V$ versus $V_h$.  
Does use of either
defintion produce relationships similiar to the equation of state the 
bulk HS fluid?   Second, what are the effects of confinement and average 
density on the transverse self-diffusivity of fluids
confined to pores narrower 
than those previously examined\cite{mittal2006}~
(i.e., $H<3.5 \sigma$)? 
Third, how do the behaviors of the confined and bulk HS fluid systems 
compare under conditions of equal activity as opposed to equal density?
Finally, does the robust relationship between excess entropy $s^{\text
  {ex}}$ (relative to ideal gas) and self-diffusivity $D$, previously
discovered for fluids confined to gap sizes larger than $H=3.5 \sigma$ in this
system,\cite{mittal2006}~continue to hold for very narrow pores
($H<3.5 \sigma$)?  
By addressing
these four questions, we can make significant headway
not only in indentifying the regions in the $H-\rho$ and $H-\xi$
planes of parameter space
where the confined HS system significantly deviates from the
bulk HS fluid, but also in probing the microscopic 
mechanisms for such deviations.  

\section{Simulation Methods \label{Method}} 
To explore these issues, we have calculated
the thermodynamic properties of confined and bulk HS fluids 
using grand canonical 
transition-matrix Monte Carlo (GC-TMMC) 
simulations,\cite{jeff1,jeff2}~and we have tracked
their single-particle dynamics via
discontinuous molecular dynamics (DMD) simulations.\cite{rap}~
To simplify the notation in this article, we have implicitly 
non-dimensionalized all quantities by appropriate
combinations of a characteristic length scale 
(which we take to be the HS particle diameter $\sigma$) and 
time scale (which we choose to be $\sigma \sqrt{m \beta}$, 
where $m$ is particle mass, $\beta=[k_{\mathrm B}T]^{-1}$, $k_{\mathrm B}$
is the Boltzmann constant, and $T$ is temperature).  As a result, all 
quantities with dimensions of energy are understood to be ``per 
$k_{\mathrm B}T$'', the only energy scale in the problem.

The DMD simulations each involved $N=1500$ 
identical HS particles. For the bulk fluid, the particle centers were contained within a cubic
simulation cell of $V_h = N/\rho_h$, and periodic
boundary conditions were applied in all three directions.  For the
confined fluid, particle centers were contained within a rectangular
parallelepiped simulation cell of $V_h = h_xh_yh_z$, where
$h_z=H-1$ and $h_x=h_y=[N/(h_z\rho_h)]^{1/2}$.  Periodic boundary 
conditions were applied in the $x$ and $y$
directions and perfectly reflecting, smooth hard walls were placed
so that particle centers were trapped in the region $0 < z < h_z$.  
The self-diffusivity $D$ of the fluid was obtained
by fitting the long-time ($t \gg 1$) behavior of the average mean-squared 
displacement of the particles 
to the Einstein relation $\left<\Delta {\bf r}_d^2\right> = 2dDt$, where
$\Delta {\bf r}_d^2$ corresponds to the mean-square displacement per particle
in the $d$ periodic directions ($d=2$,3 for the confined and bulk
fluid, respectively).  To verify that system-size effects in the
periodic
directions on $D$
were insignificant, we checked that our calculated values for $D$
for several state points compared favorably with those we
obtained using either $N=3000$ or
$N=4500$ particles.

The GC-TMMC simulations each utilized a simulation cell
of size $V_h=1000$.  For the bulk fluid, the cell was cubic with 
$h_x=h_y=h_z=10$.  For the confined fluid, the cell was a rectangular
parallelepiped with $h_z=H-1$ and $h_y=h_z=\sqrt{1000/h_z}$.  GC-TMMC 
simulations require a specified value for the
activity~$\xi$\cite{note1}~(i.e., $N$ is allowed to fluctuate), and we set $\xi=1$ 
in all cases.   
The key quantities that we extracted from the simulations were the 
normalized total particle number probability distribution $\Pi(N)$
and the $N$-specific 
spatial density distribution $\rho(N,\bf r)$, both evaluated over a 
range of particle numbers spanning from $N = 0$ to $N = 984$. 
Thermodynamic properties at other 
values of activity $\xi$ were readily obtained via the histogram 
reweighting technique\cite{swendsen}~to shift the original $\Pi(N)$ 
distribution to one representative of the particle numbers visited at the 
selected $\xi$.  We found that we obtained statistically 
indistinguishable results for systems with $V_h=500$, indicating again
that noticeable artifacts associated with system size were not present. 

By employing basic arguments from statistical mechanics,\cite{azp00,htdavis}~
one can use the equilibrium 
information from GC-TMMC simulations to compute thermodynamic
properties of interest.
Specifically, the grand potential $\Omega$ can be calculated directly 
from the normalized particle number distribution,\cite{azp00,errington2005}~
\begin{equation}
\Omega = {\text {ln}}\Pi(0).
\label {p_t}
\end{equation}
For the bulk HS fluid, we also have $V=V_h$, and thus
$\rho = \sum_N N \Pi(N)/V=\rho_h$.  
Moreover, the pressure of the bulk fluid $P$ is equal to 
the negative of the grand potential density, $P=-\Omega/V$.  
On the other hand, for the HS fluid confined between hard
walls, we have $V=V_h/(1-H^{-1})$, and thus 
$\rho = \sum_N N \Pi(N)/V=(1-H^{-1})\rho_h$.  In this case, negative 
grand potential density $-\Omega/V$ represents an average 
transverse pressure
acting parallel to the confining walls.\cite{henderson}~
In the reduced units adopted here, the component of the pressure
tensor acting normal to the walls is equal to the local fluid
density in contact with a hard wall, $P_z(N)= \rho(N,z=0.5)=\rho(N,z=H-0.5)$, a
consequence of an exact statistical mechanical sum rule for this 
system.\cite{fisher}~Finally, the molar excess 
entropy $s^{\mathrm{ex}}=S^{\mathrm{ex}}/N$ is determined using the following
expression,\cite{mittal2006,jeff2006}~
\begin{eqnarray}
S^{\mathrm{ex}}(N) = \ln [\Pi(N)/\Pi(0)]
  -N{\mathrm {ln}}{\xi} + \ln N! \nonumber \\ 
- N\ln N +\int \rho(N,{\bf r}) \ln
\rho(N,{\bf r}) d {\bf r}.
\label{s1}
\end{eqnarray}
Below, 
we describe how the above methods were employed in this study to 
characterize the behaviors of the confined and bulk HS fluids.

\section{Results and Discussion \label{Results}} 

\subsection{Volume definition and the equation of state\label{EOS-V}}
One of the most practically important and well-understood properties
of the bulk, equilibrium HS fluid is its equation of state $P(\rho)$,
which quantifies how its pressure varies with density.  For 
densities below the freezing 
transition ($\rho \lesssim 0.943$), this relationship is accurately 
described by the semi-empirical Carnahan-Starling equation
$P (\rho) \approx P_{\mathrm {CS}}(\rho) = 
\rho (1+\phi+\phi^2-\phi^3)/(1-\phi)^3$,\cite{CarSta}~where
$\phi=\pi\rho/6$ is the packing fraction of the spheres.  

Much less is known about the global behavior of the pressure
tensor for the HS fluid confined between smooth hard
walls.  
One obvious question is, do the relationships 
between the transverse and normal components of the pressure tensor
and ``average'' density (defined as either $\rho$ or $\rho_h$) show
quantitative similarities to the equation of state of the bulk fluid? 
Although the inhomogeneous
structuring of the fluid might be expected to give rise to some
nontrivial deviations from bulk fluid behavior, the main qualitative      
trends should be the same:
compressing the fluid increases the interparticle 
collision rate and, consequently, the individual 
components of the pressure tensor.
  
\begin{figure}
\scalebox{1.0}{\includegraphics{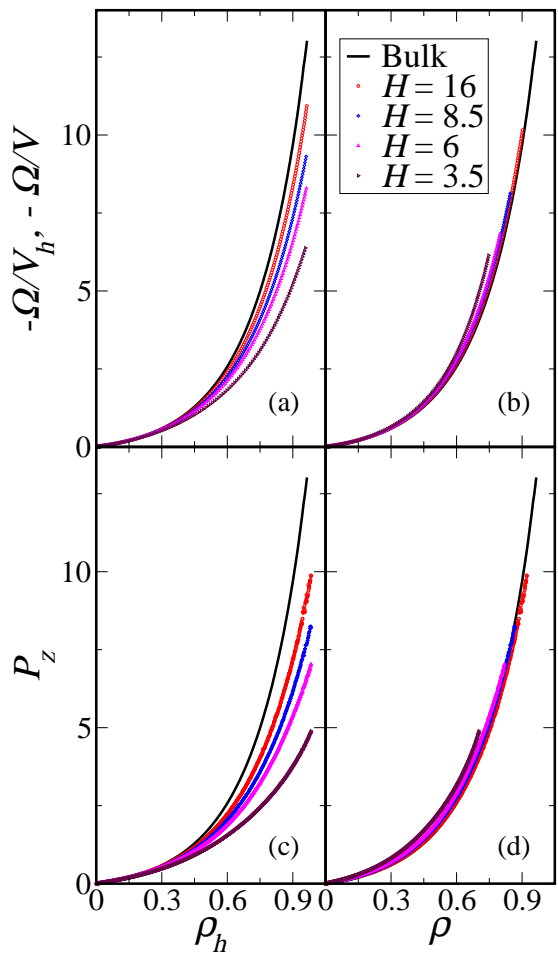}}
\caption{\label{PressureT} Equation of state of a confined HS fluid
  between hard walls separated by center-accessible distance $H-1$.  
Top panels show the negative of the 
grand potential density (average transverse pressure) 
versus average fluid density 
calculated using (a) the center-accessible volume $V_h$ and (b) the 
total
volume $V$.  Bottom panels illustrate the normal pressure versus
average density calculated using (c) the center-accessible volume $V_h$
and (d) the total volume $V$.} 
\end{figure}

In Fig.~\ref{PressureT}, we compare the bulk fluid equation of state 
to our GC-TMMC simulation data for
the average transverse and normal components of the pressure tensor. 
We focus here on confined fluids with $H=3.5$, 6, 8.5, and 16.  
In top panels (a) and (b), negative grand
potential density is plotted
versus average density, adopting the $V_h$ (center accessible) and $V$
 (total) volume conventions, 
respectively.  The density dependencies of $P_z$ 
are similarly displayed in panels (c) and (d).  Focusing on plots (a)
and (c), one finds a family of curves for $-\Omega/V_h$ and
$P_z$ that are qualitatively similar to the bulk fluid behavior, 
with the main difference being that systems
with smaller $H$ have weaker $\rho_h$ dependencies (higher apparent
compressibilities).  This difference
appears
logically consistent with the earlier observation\cite{mittal2006}~
that confined HS fluids
also have faster single-particle dynamics as compared to the corresponding 
bulk fluid with the same $\rho_h$.  

Interestingly,
the corresponding quantities plotted in (b) and (d) 
using the total volume $V$ convention approximately collapse onto a
single curve.   
This means that the $\rho$ dependencies of 
both $-\Omega/V$ and $P_z$, for each value of $H$ investigated, can 
approximately be described by the equation of state of the bulk fluid 
$P(\rho)$.  This trend also appears consistent with the approximate 
collapse of self-diffusivities for confined HS fluids onto the bulk behavior 
when plotted together on a single graph versus~$\rho$.\cite{mittal2006}~
Although there are clearly some quantitative 
deviations from bulk behavior for the smallest pores in panels (b) and
(d) of Fig.~\ref{PressureT}, we found that the 
following simple relationship can describe the $\rho$ dependence
of the grand potential density to within at least 25\% for $H \ge 3.5$:
\begin{equation}
\label{PressCS}
\Omega (\rho,H)/ V \approx - P_{\text {CS}} (\rho).
\end{equation}     
We will use this approximate 
relationship below to help construct an analytical
model for predicting the excess adsorption of fluid in a model slit
pore. 

\subsection{Interfacial free energy and excess adsorption}
Given the approximate collapse of the thermodynamic data for the
confined HS fluid when plotted against average density $\rho$, it is 
natural to ask whether there is a connection to the behavior of the 
interfacial free energy and the surface excess adsorption of the 
fluid at a single hard wall. 

The interfacial free energy of the HS fluid near a hard wall 
is defined as the excess grand potential of the fluid 
(relative to bulk) per unit fluid-wall contact area.  
Similar to average density, its numerical
value depends on the arbitrary 
choice of dividing surface,\cite{hendersonJCP,bryk}~although
different choices provide equivalent thermodynamic descriptions
of the system if applied self-consistently.  
If one chooses the
plane of closest approach of the particle centers to the wall as
the dividing surface, then the following expression yields the 
interfacial free energy:
\begin{equation}
\gamma_{h}^{\infty} = \lim_{H \to \infty} \left[\frac{\Omega(\rho,H)}{V_h} +
P_{\mathrm {b}}\right]\frac{(H-1)}{2},
\label{gamma_h}
\end{equation}
where $P_{\mathrm {b}}$ is the pressure of the bulk fluid in
equilibrium with the pore fluid.  Stated differently, $\rho$ of the pore fluid
is determined by $H$ and the requirement that it adopt the same
activity $\xi$ as the bulk HS fluid of pressure $P_{\mathrm {b}}$.  
There is an accurate approximate equation due to Henderson and
Plischke\cite{Henderson1985}~for predicting how $\gamma_{h}^{\infty}$
depends on the packing fraction of the bulk fluid $\phi_{\mathrm
  {b}}=\pi\rho_{\mathrm{b}}/6$,
\begin{equation}
\gamma_{h}^{\infty} \approx -\frac{9}{2\pi} \phi_{\mathrm
  {b}}^2 \frac{\left[1 + (44/35)\phi_{\mathrm {b}} - (4/5)
  \phi_{\mathrm {b}}^2 \right]}{(1-\phi_{\mathrm {b}})^3}.
\label{gamm_h_hend}
\end{equation}

If one instead chooses the physical surface of the wall to be
the dividing surface, then a slightly different equation emerges:
\begin{eqnarray}
\label{gamm1}
\gamma^{\infty} &=& \lim_{H \to \infty} \left[\frac{\Omega(\rho,H)}{V} +
P_{\mathrm {b}}\right]\frac{H}{2} \\
&=& \gamma_{h}^{\infty} + P_{\mathrm {b}}/2
\label{gamm2}
\end{eqnarray}
Substituting the Carnahan-Starling equation of state for 
$P_{\mathrm {b}}$ and Eq.~\ref{gamm_h_hend} for $\gamma_{h}^{\infty}$ into Eq.~\ref{gamm2} results in the following
analytical estimate for $\gamma^{\infty}$,
\begin{equation}
\gamma^{\infty} \approx \frac{3}{\pi} \phi_{\mathrm {b}} \frac{\left[1 - (1/2)\phi_{\mathrm {b}} - (31/35) \phi_{\mathrm {b}}^2 +(1/5)\phi_{\mathrm {b}}^3 \right]}{(1-\phi_{\mathrm {b}})^3}
\label{gamm_hend}
\end{equation}
Given that we have already observed that other
properties of the confined HS fluid approximately
collapse when plotted versus $\rho$ (based on total volume $V$), 
we choose to focus our attention from this point 
forward on $\gamma^{\infty}$, the
interfacial free energy that is also based on $V$.

As we demonstrated in the previous section, one can readily 
determine the quantities on the right-hand side of 
Eq.~\ref{gamm1} for finite values of $H$ using GC-TMMC simulations.
As a result, these simulations might also provide a reasonably 
accurate means for estimating $\gamma^{\infty}$, assuming that $H$
can be chosen large enough so that the perturbations to the fluid 
caused by the two confining walls do not significantly 
interfere with one another
(i.e., so that so-called ``finite-size'' or frustration 
effects of confinement do
not occur).  Although, it is not clear {\em a priori} how large $H$ must be to
achieve this, one might reasonably expect that the pore would need to
be at least several particle diameters in width.

\begin{figure}
\scalebox{1.00}{\includegraphics{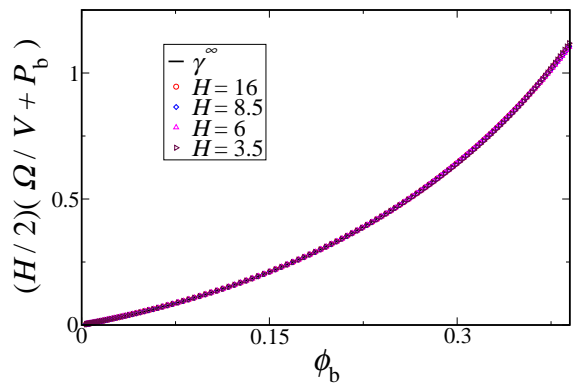}}
\caption{\label{Act-G_H} 
The quantity $(H/2)\left[\Omega/V + P_{\mathrm {b}}\right]$ calculated
from our GC-TMMC simulations for various $H$ 
along with the $H \to \infty$ limit, $\gamma^{\infty}$, computed using 
Eq.~\ref{gamm_hend}.  Data are plotted
as a function of  $\phi_{\mathrm b}$, the 
packing fraction of the bulk HS fluid 
that is in equilibrium with the pore fluid.} 
\end{figure}

As a test of this idea, we present in Fig.~\ref{Act-G_H} values of the quantity
  $\left[\Omega(\rho,H)/V + P_{\mathrm {b}}\right](H/2)$ calculated
from our GC-TMMC simulations for various $H$ 
along with the single-wall quantity
$\gamma^{\infty}$ of Eq.~\ref{gamm_hend}, which is the $H \to \infty$
  limit.  
All data are plotted
as a function of bulk packing fraction $\phi_{\mathrm b}$.  
Interestingly, the plot reveals
that the simulated curves for $H \ge 3.5$ all collapse, to within
an excellent approximation, 
onto that for $\gamma^{\infty}$.  In other words,
\begin{equation}
\gamma^{\infty} \approx \left[\frac{\Omega(\rho,H)}{V} +
P_{\mathrm {b}}\right]\frac{H}{2}
\label{approx_gamm}
\end{equation}
independent of $H$ for $H \ge 3.5$. 
This implies
that single-wall behavior such as $\gamma^{\infty}$ 
can be estimated with great accuracy in this system 
from a simulated slit-pore of width $H=3.5$, which 
can only accomodate a fluid film three particle layers thick.  
This result, while very robust, is 
somewhat surprising because the single-wall 
density profiles decay slowly enough to expect appreciable
interference or frustration 
effects at pore sizes as small as $H=3.5$. 
However, similar to the picture that emerged from the behavior of the 
equation of state in the previous section, any interference
that does occur apparently cancels in determining
the average properties of the confined HS fluid, which remain
remarkably ``bulk-like'' even for these very thin films.  

So, when do interference effects due to packing frustration of
the wall-induced particle layers begin to occur?
We can probe this issue by taking the analysis one step further.
Specifically, if one uses Eq.~\ref{PressCS} to substitute for
$\Omega(\rho,H)/V$ in Eq.~\ref{approx_gamm}, differentiates both
sides of Eq.~\ref{approx_gamm} with repect to chemical potential, and
invokes the Gibbs adsorption equation $\partial \gamma^{\infty}/
\partial \mu = - \Gamma^{\infty}$, then upon rearranging 
one arrives at the following
simple equation for predicting the pore density~$\rho$:
\begin{equation}
\rho \approx \rho_{\mathrm b} + \frac{2 \Gamma^{\infty}}{H}
\label{adsorption_eq}
\end{equation}   
The quantity $\Gamma^{\infty}$ is the standard surface
excess density for a HS fluid next a to {\em single} hard wall, and, within
the above approximations, it is given by  
\begin{equation}
\Gamma^{\infty} = -\frac{3 \phi_{\mathrm {b}} \left[1+ a_1 \phi_{\mathrm
      {b}} +a_2 \phi_{\mathrm
      {b}}^2 + a_3 \phi_{\mathrm
      {b}}^3 +a_4 \phi_{\mathrm
      {b}}^4 \right]}{\pi (1+4 \phi_{\mathrm
      {b}} + 4 \phi_{\mathrm
      {b}}^2 - 4\phi_{\mathrm
      {b}}^3 +\phi_{\mathrm
      {b}}^4)},
\label{gamma_II}
\end{equation} 
where $a_1 = 1$, $a_2 = -221/70$, $a_3 = 4/5$, and $a_4 = -1/5$. 

\begin{figure}
\scalebox{1.00}{\includegraphics{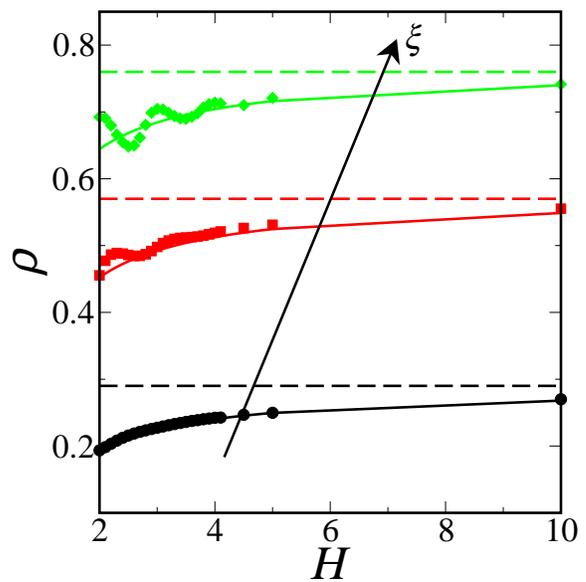}}
\caption{\label{Act-R_H} Pore fluid density $\rho$ 
as a function of pore width
$H$ at different values of activity [$\ln \xi=0.4$, 4.4, and 8.8]. 
Filled symbols and solid lines correspond to the GC-TMMC data and the 
predictions of Eq.~\ref{adsorption_eq}, respectively. 
The dashed lines correspond to the 
bulk density $\rho_{\text b}$ for a given activity $\xi$.} 
\end{figure}

Fig.~\ref{Act-R_H} shows the predictions of 
the simple analytical model of Eq.~\ref{adsorption_eq} and~\ref{gamma_II}
compared to the simulated pore density~$\rho$
as a function of~$H$.
>From the plot, it is evident that the average pore density can be predicted
based on knowledge of only the single-wall surface excess
$\Gamma^{\infty}$ unless the
fluid is {\em both} dense {\em and} confined to pores narrower than approximately
three particle diameters.  Under those restrictive 
conditions, the single-wall model misses the emergence of oscillations
in the pore density.  These oscillations cannot be solely attributed
to single-wall ``layering'' in the density profile because pronounced layering
also occurs for dense fluids with $H\gg3$, where the analytical 
model is still very
accurate.  Rather, the oscillations must be due to packing frustration
associated with the interference of the layers emerging from the two
confining walls, 
which apparently becomes significant in this system 
only for very narrow pores and high fluid density.

\subsection{Comparing bulk and pore fluid self-diffusivities}

The last two sections demonstrated that some of the
average thermodynamic properties of the confined HS fluid are 
very similar to those of the bulk fluid if the two systems are
compared at equal values of the average density $\rho$ (based on the
total system volume).  Deviations occur only when the fluid is both
dense and confined to pores narrower than approximately three particle
diameters.  In a previous study,\cite{mittal2006}~we have also shown
that the self-diffusivity $D$ of the confined HS fluid is
approximately equal to that of 
the bulk fluid with the same $\rho$ 
for $H>3.5$ over a fairly broad range of $\rho$.
Here, we carefully investigate the $H$-dependency 
of pore self-diffusivity at constant $\rho$ for narrow pores, with a focus on 
understanding when packing frustration causes the correlation between
$D$ and $\rho$ to break down.  We also investigate the $H$-dependency of
$D$ for the confined fluid under the constraint of constant imposed activity
$\xi$.  We find that this latter behavior can be essentially 
predicted in advance,
given the known connection between $D$ and $\rho$\cite{mittal2006}~
and the ability to predict $\rho$ from $\xi$ and $H$ discussed in the
previous section.      

\begin{figure}
\scalebox{1.}{\includegraphics{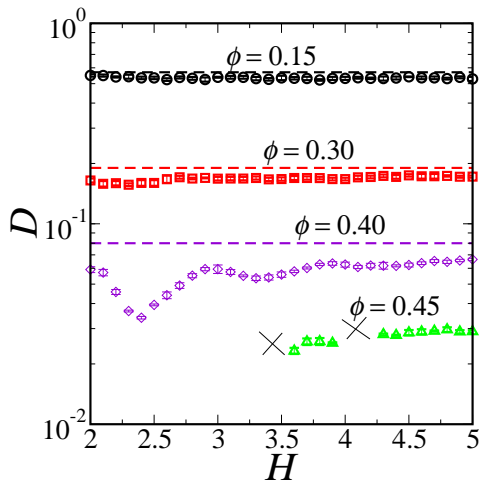}}
\caption{\label{Phi-D_H} Self-diffusivity $D$ as a function of pore width 
$H$ at different pore fluid packing fractions $\phi$. 
For $\phi=0.45$, crosses mark regions for which the
confined system penetrates into the
fluid-solid coexistence region or the solid phase region on its 
equilibrium phase diagram.\cite{marjolein2006}}
\end{figure}

We begin here by examining how $H$ affects $D$ at constant $\rho$ 
using the DMD simulations described earlier.  Specifically, we
plot in Fig.~\ref{Phi-D_H} the self-diffusivity $D$ of the 
bulk and confined HS fluid for $H=2$ to 5 and various pore packing 
fractions ($\phi\equiv\pi \rho/6=0.15,$ 0.30, 0.40, and 0.45).  
What is plainly evident is 
that up to fairly dense packing fractions ($\phi<0.40$), 
$D$ of the confined fluid shows no significant 
deviations from bulk behavior (dashed line) even when in very restrictive 
pores (e.g. $H=2$).  In fact, quantitative deviations are
prominent ($> 25 \%$) only in the high density $(\phi \ge 0.4)$ and
small pore ($H<3$) limit.  Note that an equilibrium 
fluid at $\phi=0.45$ cannot be accessed over the full 
$H$ range because the system penetrates into the
fluid-solid coexistence region or the solid phase region on its 
phase diagram.\cite{marjolein2006}~

\begin{figure}
\scalebox{1.00}{\includegraphics{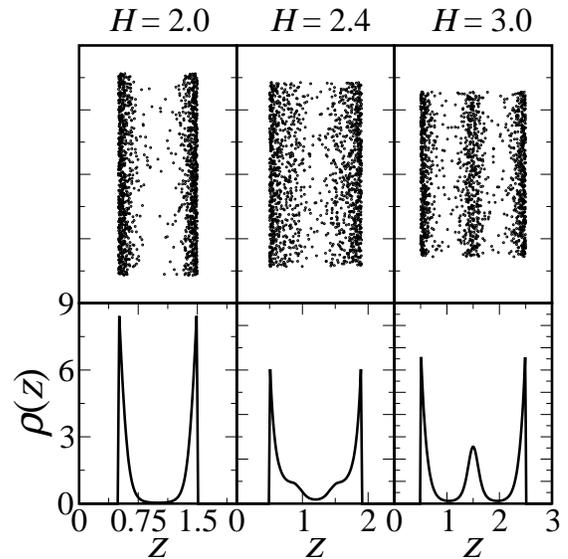}}
\caption{\label{Profile} 2D projections of typical instantaneous 
particle configurations of the confined HS fluid are shown
(top) along with equilibrium density profiles 
$\rho(z)$ (bottom), where $z$ represents the positional 
coordinate normal to the walls.}
\end{figure}

To gain a more physical understanding of the variations in $D$ at 
constant $\phi$ that occur under conditions of high $\phi$ and low $H$, 
we plot in Fig.~\ref{Profile} 
the 2D projections of instantaneous particle configurations 
of the confined HS fluid for 
$H=2.0$, 2.4, and 3.0 at $\phi=0.40$, state points that show very
different dynamical behaviors.  We also present the
corresponding density profiles $\rho(z)$ normal to the walls.
This figure shows well-developed layering structures for both $H=2.0$ (two
particle layers) and $H=3.0$ (three particle layers).  However, the
system at $H=2.4$ shows considerably more
packing frustration.  In particular, the individual density peaks 
are reduced in this case 
because the spacing is such that it is ``in between'' distances that 
naturally accomodate 
either two or three layers.  The pore diffusivity 
is also lowest for $\phi=0.4$ at $H=2.4$ as shown in 
Fig.~\ref{Phi-D_H}.  Similar oscillations in $D$, which are much smaller in
magnitude and decay with increasing $H$, occur at larger separations 
with the minima again coinciding with spacings that do not naturally 
accommodate an integer number
of particle layers.  In short, for small 
enough pores and high enough densities, the frustrated layering
of particles normal to the confining walls 
significantly slows down the single-particle dynamics in the direction
parallel to the walls.      

The trend that increased layering leads to faster dynamics 
may initially appear counter-intuitive, especially if one
tries to understand it by drawing an analogy with the bulk HS system.  
In the bulk HS system, compressing the fluid increases the structural
ordering\cite{torquato2000,truskett2000}~but {\em reduces} 
the self-diffusivity.  In contrast, as can be clearly seen in 
Fig.~\ref{Phi-D_H} and \ref{Profile}, 
increased layering in the normal direction 
(i.e., less uniform
density profiles) correlates with faster dynamics.  However, these two represent
fundamentally different systems undergoing different changes.  
In the bulk HS system, increasing the density not only increases the 
structural order, but it also reduces the entropy (or average free volume) of 
the particles in the fluid.  This compression-induced 
reduction in free volume is not
surprisingly correlated with slower dynamics.
However, the confined HS system actually maximizes its entropy 
(or average free volume) at fixed average density and $H$ by adopting an
inhomogeneous density profile with pronounced layering.\cite{kjellander}~ 
Our results show that, for constant $\rho$, 
the values of (small) $H$ that frustrate the ability of the system to 
form an integral 
number of particle layers also tend to reduce the 
single-particle mobility in the direction parallel to the walls.  We
return to investigate the potential connection between dynamics and entropy
of the confined HS fluid in the next section.   

Another important point concerning 
the frustration-induced oscillations 
in $D$ of Fig.~\ref{Phi-D_H} is that they are distinct from the
oscillations in $D$ that occur as a function of $H$ at constant
activity $\xi$.\cite{ted1985,ted1987}~
The latter are inevitably impacted by oscillations in average pore 
density, whereas the average density is being
controlled for (held constant) in Fig.~\ref{Phi-D_H}.  
Deviations from bulk behavior at fixed average density 
are purely frustration-induced {\em finite-size 
effects}, and the relative importance of these types of
deviations has been a long-standing question in the 
study of confined fluids.\cite{mckenna}~ 

Interestingly, if one
compares the locations of the oscillations in $D$ versus $H$ at $\phi=0.40$
in Fig.~\ref{Phi-D_H} with 
the fluid-solid phase boundary of this system presented by 
Fortini {\em et al.},\cite{marjolein2006}~one also finds a strong correlation
between slow dynamics and proximity of the fluid 
to the phase boundary.  In other
words, the same packing frustration that is giving rise to slow
dynamics also appears to ultimately 
promote the formation of an ordered solid phase.  This argues that 
the effect of confinement on the phase diagram of the
system can provide important insights into 
how confinement
impacts single-particle dynamics.
The consequences of this could be significant for the
strategies that are typically employed to study supercooled
and confined liquids.
For example, weak polydispersity is commonly incorporated into
model fluid systems in order to study them under conditions where
the corresponding 
monatomic fluid would rapidly crystallize.  A cautionary note that
follows from the above discussion is that one should not readily 
assume that the behaviors of the polydisperse and monatomic systems 
are trivially related,
and that the former only differs from the latter 
in that its liquid state is
kinetically accessible over a broader range of conditions. 
The phase diagrams of polydisperse materials 
are considerably more complex than monatomic
systems (even in the bulk.\cite{sollich})~Thus, one should expect
confinement to impact the
dynamics of polydisperse systems 
in ways that are not easily relatable to the behavior of the
corresponding monatomic fluids. 

\begin{figure}
\scalebox{1.00}{\includegraphics{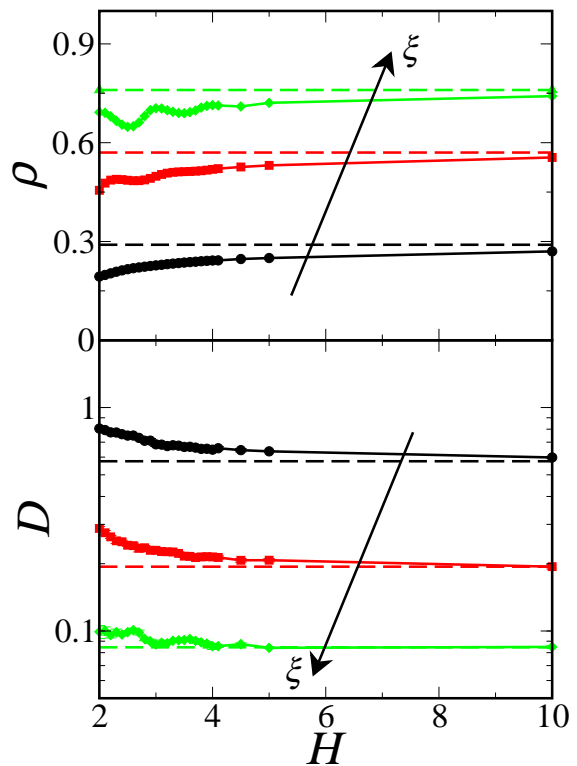}}
\caption{Average density $\rho$ and self-diffusivity $D$ as 
a function of pore size $H$ for the confined HS fluid 
in equilibrium with the bulk HS fluid at 
at a given activity $\xi$ [$\ln \xi=0.4$, 4.4, and 8.8].
Dashed lines correspond to the density $\rho_b$ and self-diffusivity 
$D$ of the bulk HS fluid at the given activity $\xi$. 
\label{Act-R_D}}
\end{figure}

We now turn our attention to 
the $H$-dependent diffusivity behavior of the confined HS fluid at
fixed activity $\xi$ (i.e., in chemical equilibrium with the bulk). 
As can be ascertained from the strong correspondence between
$D$ and $\rho$ in Fig.~\ref{Phi-D_H}, the dynamical behavior at constant
$\xi$ can be largely predicted 
in advance if one simply has knowledge  of 
how $H$ influences $\rho$ at constant $\xi$ (e.g., from simulation or the
analytical model of Eq.~\ref{adsorption_eq} and~\ref{gamma_II}).
In Fig.~\ref{Act-R_D}, we provide the $H$-dependent data along
constant $\xi$ paths for the quantities $\rho$
and $D$ determined from GC-TMMC and
DMD simulations, respectively.  One initial observation is that $\rho$
is always less than $\rho_b$ for finite $H$, and, 
as should be expected based on this, 
$D$ is larger in the pores than in the equilibrium 
bulk fluid.  Note that this type of
physically-intuitive connection between average density and dynamics 
would be completely lost, however, if one instead chooses 
$\rho_h$ as the definition for average density, which is
significantly greater than $\rho_b$ for finite $H$.  More generally, the
reliability of approximate theories for transport properties
in inhomogeneous fluids could be particularly sensitive to how averaging
is handled, which might help to explain why an earlier 
kinetic theory\cite{ted1987}~predicts that confining a fluid
at constant $\xi$ significantly decreases $D$, the 
opposite of what is seen in the MD simulation
data of Fig.~\ref{Act-R_D}. 

A second observation about the data in Fig.~\ref{Act-R_D} 
is that there are negative oscillatory 
deviations in $\rho$ (relative to bulk) with $H$ at high $\xi$ in
small pores, which one might expect to produce similar positive
oscillations in $D$.   However, the frustration-induced
negative deviations from bulk behavior in the $D$ versus $H$ relationship at
constant $\rho$ shown in Fig.~~\ref{Phi-D_H} appear
to largely cancel this effect.  The net result is that $D$ is
strikingly similar to bulk behavior, even for small $H$, along paths
of constant (and sufficiently high) $\xi$.  

\subsection{Diffusivity and excess entropy}
The oscillatory 
data in Fig.~\ref{Phi-D_H} clearly show that average density alone
cannot predict the self-diffusivity of the HS fluid if the 
fluid is both dense and confined to a pore smaller than approximately
three diameters.  Is there another thermodynamic quantity that can
predict diffusivity behavior is these narrow pores?  One promising
candidate is the excess entropy $s^{\text {ex}}$ (relative to ideal
gas), which recent DMD simulations\cite{mittal2006}~demostrate, 
to an excellent
approximation, determines the self-diffusivity of the HS fluid
confined between hard walls for $H>3.5$.  Here, we explore its
relationship to self-diffusivity in smaller pores. 
\begin{figure}
\scalebox{1.00}{\includegraphics{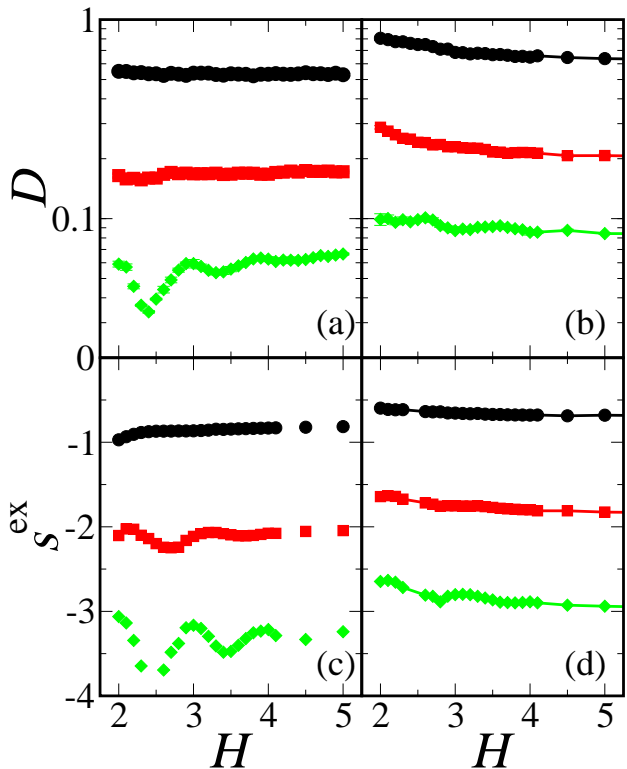}}
\caption{(a) Self-diffusivity $D$ and (b) excess entropy $s^{\text {ex}}$ 
as a function of pore size $H$ for the confined HS fluid 
at a given pore packing fraction $\phi=\pi\rho/6$. 
Data from top to bottom correspond to $\phi=0.15$, 0.3, and 0.4.
(c) Self-diffusivity $D$ and (d) excess entropy $s^{\text {ex}}$ as 
a function of pore size $H$ for the confined HS fluid 
at a given activity $\xi$. 
Data from top to bottom correspond to $\ln \xi=0.4$, 4.4, and 8.8.
\label{Phi-Act-D_s}}
\end{figure}

Fig~\ref{Phi-Act-D_s} shows the data for $D$ and $s^{\text {ex}}$
of the confined fluid collected from our DMD and GC-TMMC simulations 
[data at fixed pore 
packing fraction $\phi$ provided in panels (a) and (c), and 
data at fixed activity $\xi$ provided in (b) and (d)].
Irrespective of the thermodynamic path, strong qualitative 
correspondence is observed between $D$ and $s^{\text {ex}}$, including
the ``in-phase'' oscillations that emerge at small $H$.  In other
words, self-diffusivity and excess entropy appear to be affected
in a very similar way by confinement, even for the very narrow pores.

\begin{figure}
\scalebox{1.00}{\includegraphics{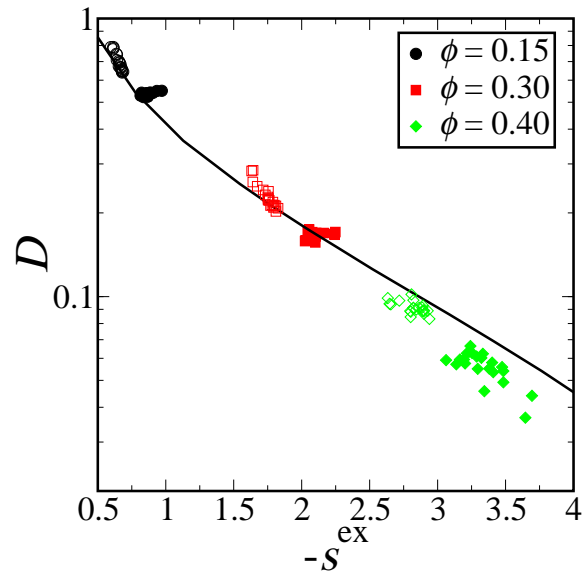}}
\caption{Self-diffusivity $D$ vs negative excess entropy 
per particle -$s^{\text {ex}}$ for the bulk 
HS fluid (solid curve) and for the HS fluid confined 
between smooth hard walls (symbols).  The filled symbols correspond 
to the confined system at a fixed pore packing fraction $\phi$ as shown 
on the legend and empty symbols are for system at a given activity 
$\xi$ for which bulk packing fraction is given on the legend. 
\label{D-s}}
\end{figure}

To scrutinize the quantitative accuracy of the relation between
the two variables, we also plot all data corresponding to 
constant pore packing fraction $\phi$ (filled symbols) and 
constant $\xi$ (empty symbols) paths in Fig~\ref{D-s} in the 
$D$-$s^{\text
  {ex}}$ plane.  As can be seen, most of the data falls very close the 
curve for the bulk HS fluid, indicating that excess entropy (a static 
quantity) can indeed approximately predict the implications of confinement
for self-diffusivity.  The largest deviations are for the fluid that
has the highest pore packing fraction of $\phi=0.4$.  

This data is yet 
one more manifestation of a larger trend seen throughout this paper.  Namely,
that the confined HS fluid, by measure of many of its average properties,
has behavior very similar to that of the bulk fluid.  It changes
character only under a fairly restrictive set of conditions, when the
pore fluid is dense ($\phi \ge 0.4$) and when it is confined to pores
smaller than approximately three particle diameters in width.

\section{Conclusions}
In conclusion, we have presented new comprehensive simulation 
results for the HS fluid confined between smooth 
hard walls. The results elucidate thermodynamic 
and dynamic behavior of this system over a wide range of 
system conditions, further clarifying the precise role of 
confinement on average fluid properties and the most useful
way to define average density for this system.  
One perhaps unexpected 
result is that, for most conditions, 
the average behavior of the confined HS fluid 
is very similar to that the bulk fluid.  
Frustration-induced finite effects do emerge in this system, but they are
only prominent for very small pores (dimensions smaller 
than approximately three 
particle diameters) and high fluid densities where the system 
approaches the confinement-shifted fluid-solid phase boundary. 

\begin{acknowledgments}
JM acknowledges the financial support from a Continuing University Fellowship 
of The University of Texas at Austin.
TMT and JRE acknowledge the financial support of the National
Science Foundation Grants No. CTS-0448721 and CTS-028772,
respectively, and the Donors of the American Chemical Society
Petroleum Research Fund Grants No. 41432-G5 and 43452-AC5, respectively.
TMT also acknowledges the support of the David and Lucile Packard
Foundation and the Alfred P. Sloan Foundation.
The Texas Advanced Computing Center (TACC) and University at Buffalo 
Center for Computational Research provided computational 
resources for this study.
\end{acknowledgments}


\begin{thebibliography}{32}
\expandafter\ifx\csname natexlab\endcsname\relax\def\natexlab#1{#1}\fi
\expandafter\ifx\csname bibnamefont\endcsname\relax
  \def\bibnamefont#1{#1}\fi
\expandafter\ifx\csname bibfnamefont\endcsname\relax
  \def\bibfnamefont#1{#1}\fi
\expandafter\ifx\csname citenamefont\endcsname\relax
  \def\citenamefont#1{#1}\fi
\expandafter\ifx\csname url\endcsname\relax
  \def\url#1{\texttt{#1}}\fi
\expandafter\ifx\csname urlprefix\endcsname\relax\def\urlprefix{URL }\fi
\providecommand{\bibinfo}[2]{#2}
\providecommand{\eprint}[2][]{\url{#2}}

\bibitem[{\citenamefont{Davis}(1996)}]{htdavis}
\bibinfo{author}{\bibfnamefont{H.~T.} \bibnamefont{Davis}},
  \emph{\bibinfo{title}{Statistical Mechanics of Phases, Interfaces, and Thin
  Films}} (\bibinfo{publisher}{VCH}, \bibinfo{year}{1996}).

\bibitem[{\citenamefont{Fortini and Dijkstra}(2006)}]{marjolein2006}
\bibinfo{author}{\bibfnamefont{A.}~\bibnamefont{Fortini}} \bibnamefont{and}
  \bibinfo{author}{\bibfnamefont{M.}~\bibnamefont{Dijkstra}},
  \bibinfo{journal}{J. \ Phys.: \ Condens. \ Matter}
  \textbf{\bibinfo{volume}{18}}, \bibinfo{pages}{L371} (\bibinfo{year}{2006}).

\bibitem[{\citenamefont{Thompson et~al.}(1992)\citenamefont{Thompson, Grest,
  and Robbins}}]{robbins1992}
\bibinfo{author}{\bibfnamefont{P.~A.} \bibnamefont{Thompson}},
  \bibinfo{author}{\bibfnamefont{G.~S.} \bibnamefont{Grest}}, \bibnamefont{and}
  \bibinfo{author}{\bibfnamefont{M.~O.} \bibnamefont{Robbins}},
  \bibinfo{journal}{Phys. \ Rev. \ Lett.} \textbf{\bibinfo{volume}{68}},
  \bibinfo{pages}{3448} (\bibinfo{year}{1992}).

\bibitem[{\citenamefont{Schmidt and L{\"{o}}wen}(1996)}]{matthias1996}
\bibinfo{author}{\bibfnamefont{M.}~\bibnamefont{Schmidt}} \bibnamefont{and}
  \bibinfo{author}{\bibfnamefont{H.}~\bibnamefont{L{\"{o}}wen}},
  \bibinfo{journal}{Phys. \ Rev. \ Lett.} \textbf{\bibinfo{volume}{76}},
  \bibinfo{pages}{4552} (\bibinfo{year}{1996}).

\bibitem[{\citenamefont{Zangi and Rice}(1998)}]{rice1998}
\bibinfo{author}{\bibfnamefont{R.}~\bibnamefont{Zangi}} \bibnamefont{and}
  \bibinfo{author}{\bibfnamefont{S.~A.} \bibnamefont{Rice}},
  \bibinfo{journal}{Phys. \ Rev. \ E} \textbf{\bibinfo{volume}{58}},
  \bibinfo{pages}{7529} (\bibinfo{year}{1998}).

\bibitem[{\citenamefont{Schmidt and L{\"{o}}wen}(1997)}]{lowen2dp}
\bibinfo{author}{\bibfnamefont{M.}~\bibnamefont{Schmidt}} \bibnamefont{and}
  \bibinfo{author}{\bibfnamefont{H.}~\bibnamefont{L{\"{o}}wen}},
  \bibinfo{journal}{Phys. \ Rev. \ E} \textbf{\bibinfo{volume}{55}},
  \bibinfo{pages}{7228} (\bibinfo{year}{1997}).

\bibitem[{\citenamefont{Mittal et~al.}(2006)\citenamefont{Mittal, Errington,
  and Truskett}}]{mittal2006}
\bibinfo{author}{\bibfnamefont{J.}~\bibnamefont{Mittal}},
  \bibinfo{author}{\bibfnamefont{J.~R.} \bibnamefont{Errington}},
  \bibnamefont{and} \bibinfo{author}{\bibfnamefont{T.~M.}
  \bibnamefont{Truskett}}, \bibinfo{journal}{Phys. \ Rev. \ Lett.}
  \textbf{\bibinfo{volume}{96}}, \bibinfo{pages}{177804}
  (\bibinfo{year}{2006}).

\bibitem[{\citenamefont{Dijkstra}(2004)}]{marjolein2004}
\bibinfo{author}{\bibfnamefont{M.}~\bibnamefont{Dijkstra}},
  \bibinfo{journal}{Phys. \ Rev. \ Lett.} \textbf{\bibinfo{volume}{93}},
  \bibinfo{pages}{108303} (\bibinfo{year}{2004}).

\bibitem[{\citenamefont{Auer and Frenkel}(2003)}]{auer2003}
\bibinfo{author}{\bibfnamefont{S.}~\bibnamefont{Auer}} \bibnamefont{and}
  \bibinfo{author}{\bibfnamefont{D.}~\bibnamefont{Frenkel}},
  \bibinfo{journal}{Phys. \ Rev. \ Lett.} \textbf{\bibinfo{volume}{91}},
  \bibinfo{pages}{015703} (\bibinfo{year}{2003}).

\bibitem[{\citenamefont{Kegel}(2001)}]{kegel2001}
\bibinfo{author}{\bibfnamefont{W.~K.} \bibnamefont{Kegel}},
  \bibinfo{journal}{J. \ Chem. \ Phys.} \textbf{\bibinfo{volume}{115}},
  \bibinfo{pages}{6538} (\bibinfo{year}{2001}).

\bibitem[{\citenamefont{Heni and L{\"{o}}wen}(1999)}]{heni1999}
\bibinfo{author}{\bibfnamefont{M.}~\bibnamefont{Heni}} \bibnamefont{and}
  \bibinfo{author}{\bibfnamefont{H.}~\bibnamefont{L{\"{o}}wen}},
  \bibinfo{journal}{Phys. \ Rev. \ E} \textbf{\bibinfo{volume}{60}},
  \bibinfo{pages}{7057} (\bibinfo{year}{1999}).

\bibitem[{\citenamefont{Magda et~al.}(1985)\citenamefont{Magda, Tirrell, and
  Davis}}]{ted1985}
\bibinfo{author}{\bibfnamefont{J.~J.} \bibnamefont{Magda}},
  \bibinfo{author}{\bibfnamefont{M.~V.} \bibnamefont{Tirrell}},
  \bibnamefont{and} \bibinfo{author}{\bibfnamefont{H.~T.} \bibnamefont{Davis}},
  \bibinfo{journal}{J. \ Chem. \ Phys.} \textbf{\bibinfo{volume}{83}},
  \bibinfo{pages}{1888} (\bibinfo{year}{1985}).

\bibitem[{\citenamefont{Vanderlick and Davis}(1987)}]{ted1987}
\bibinfo{author}{\bibfnamefont{T.~K.} \bibnamefont{Vanderlick}}
  \bibnamefont{and} \bibinfo{author}{\bibfnamefont{H.~T.} \bibnamefont{Davis}},
  \bibinfo{journal}{J. \ Chem. \ Phys.} \textbf{\bibinfo{volume}{87}},
  \bibinfo{pages}{1791} (\bibinfo{year}{1987}).

\bibitem[{\citenamefont{Errington}(2003{\natexlab{a}})}]{jeff1}
\bibinfo{author}{\bibfnamefont{J.~R.} \bibnamefont{Errington}},
  \bibinfo{journal}{J. \ Chem. \ Phys.} \textbf{\bibinfo{volume}{118}},
  \bibinfo{pages}{9915} (\bibinfo{year}{2003}{\natexlab{a}}).

\bibitem[{\citenamefont{Errington}(2003{\natexlab{b}})}]{jeff2}
\bibinfo{author}{\bibfnamefont{J.~R.} \bibnamefont{Errington}},
  \bibinfo{journal}{Phys. \ Rev. \ E} \textbf{\bibinfo{volume}{67}},
  \bibinfo{pages}{012102} (\bibinfo{year}{2003}{\natexlab{b}}).

\bibitem[{\citenamefont{Rapaport}(2004)}]{rap}
\bibinfo{author}{\bibfnamefont{D.~C.} \bibnamefont{Rapaport}},
  \emph{\bibinfo{title}{The Art of Molecular Dynamics Simulation}}
  (\bibinfo{publisher}{Cambridge University Press}, \bibinfo{year}{2004}),
  \bibinfo{edition}{2nd} ed.

\bibitem[{not()}]{note1}
\bibinfo{note}{The activity is defined as $\xi=\exp(\beta \mu)/\Lambda^3$,
  where $\mu$ is the chemical potential and $\Lambda$ is the de Broglie
  wavelength.}

\bibitem[{\citenamefont{Ferrenberg and Swendsen}(1988)}]{swendsen}
\bibinfo{author}{\bibfnamefont{A.~M.} \bibnamefont{Ferrenberg}}
  \bibnamefont{and} \bibinfo{author}{\bibfnamefont{R.~H.}
  \bibnamefont{Swendsen}}, \bibinfo{journal}{Phys. \ Rev. \ Lett.}
  \textbf{\bibinfo{volume}{61}}, \bibinfo{pages}{2635} (\bibinfo{year}{1988}).

\bibitem[{\citenamefont{Panagiotopoulos}(2000)}]{azp00}
\bibinfo{author}{\bibfnamefont{A.~Z.} \bibnamefont{Panagiotopoulos}},
  \bibinfo{journal}{J. Phys.: Condens. Matter} \textbf{\bibinfo{volume}{12}},
  \bibinfo{pages}{R25} (\bibinfo{year}{2000}).

\bibitem[{\citenamefont{Errington and Shen}(2005)}]{errington2005}
\bibinfo{author}{\bibfnamefont{J.~R.} \bibnamefont{Errington}}
  \bibnamefont{and} \bibinfo{author}{\bibfnamefont{V.~K.} \bibnamefont{Shen}},
  \bibinfo{journal}{J. \ Chem. \ Phys.} \textbf{\bibinfo{volume}{123}},
  \bibinfo{pages}{164103} (\bibinfo{year}{2005}).

\bibitem[{\citenamefont{Henderson and van Swol}(1984)}]{henderson}
\bibinfo{author}{\bibfnamefont{J.~R.} \bibnamefont{Henderson}}
  \bibnamefont{and} \bibinfo{author}{\bibfnamefont{F.}~\bibnamefont{van Swol}},
  \bibinfo{journal}{Mol. \ Phys.} \textbf{\bibinfo{volume}{51}},
  \bibinfo{pages}{991} (\bibinfo{year}{1984}).

\bibitem[{\citenamefont{Fisher}(1964)}]{fisher}
\bibinfo{author}{\bibfnamefont{I.~Z.} \bibnamefont{Fisher}},
  \emph{\bibinfo{title}{Statistical Theory of Liquids}}
  (\bibinfo{publisher}{The University of Chicago Press}, \bibinfo{year}{1964}).

\bibitem[{\citenamefont{Errington et~al.}(2006)\citenamefont{Errington,
  Truskett, and Mittal}}]{jeff2006}
\bibinfo{author}{\bibfnamefont{J.~R.} \bibnamefont{Errington}},
  \bibinfo{author}{\bibfnamefont{T.~M.} \bibnamefont{Truskett}},
  \bibnamefont{and} \bibinfo{author}{\bibfnamefont{J.}~\bibnamefont{Mittal}},
  \bibinfo{journal}{J. \ Chem. \ Phys.} \textbf{\bibinfo{volume}{125}},
  \bibinfo{pages}{244502} (\bibinfo{year}{2006}).

\bibitem[{\citenamefont{Carnahan and Starling}(1969)}]{CarSta}
\bibinfo{author}{\bibfnamefont{N.~F.} \bibnamefont{Carnahan}} \bibnamefont{and}
  \bibinfo{author}{\bibfnamefont{K.~E.} \bibnamefont{Starling}},
  \bibinfo{journal}{J. \ Chem. \ Phys.} \textbf{\bibinfo{volume}{51}},
  \bibinfo{pages}{635} (\bibinfo{year}{1969}).

\bibitem[{\citenamefont{Henderson}(2002)}]{hendersonJCP}
\bibinfo{author}{\bibfnamefont{J.~R.} \bibnamefont{Henderson}},
  \bibinfo{journal}{J. \ Chem. \ Phys.} \textbf{\bibinfo{volume}{116}},
  \bibinfo{pages}{5039} (\bibinfo{year}{2002}).

\bibitem[{\citenamefont{Bryk et~al.}(2003)\citenamefont{Bryk, Roth, Mecke, and
  Dietrich}}]{bryk}
\bibinfo{author}{\bibfnamefont{P.}~\bibnamefont{Bryk}},
  \bibinfo{author}{\bibfnamefont{R.}~\bibnamefont{Roth}},
  \bibinfo{author}{\bibfnamefont{K.~R.} \bibnamefont{Mecke}}, \bibnamefont{and}
  \bibinfo{author}{\bibfnamefont{S.}~\bibnamefont{Dietrich}},
  \bibinfo{journal}{Phys. \ Rev. \ E} \textbf{\bibinfo{volume}{68}},
  \bibinfo{pages}{031602} (\bibinfo{year}{2003}).

\bibitem[{\citenamefont{Henderson and Plischke}(1985)}]{Henderson1985}
\bibinfo{author}{\bibfnamefont{D.}~\bibnamefont{Henderson}} \bibnamefont{and}
  \bibinfo{author}{\bibfnamefont{M.}~\bibnamefont{Plischke}},
  \bibinfo{journal}{Proc. \ R. \ Soc. \ Lond. \ A}
  \textbf{\bibinfo{volume}{400}}, \bibinfo{pages}{163} (\bibinfo{year}{1985}).

\bibitem[{\citenamefont{Torquato et~al.}(2000)\citenamefont{Torquato, Truskett,
  and Debenedetti}}]{torquato2000}
\bibinfo{author}{\bibfnamefont{S.}~\bibnamefont{Torquato}},
  \bibinfo{author}{\bibfnamefont{T.~M.} \bibnamefont{Truskett}},
  \bibnamefont{and} \bibinfo{author}{\bibfnamefont{P.~G.}
  \bibnamefont{Debenedetti}}, \bibinfo{journal}{Phys. \ Rev. \ Lett.}
  \textbf{\bibinfo{volume}{84}}, \bibinfo{pages}{2064} (\bibinfo{year}{2000}).

\bibitem[{\citenamefont{Truskett et~al.}(2000)\citenamefont{Truskett, Torquato,
  and Debenedetti}}]{truskett2000}
\bibinfo{author}{\bibfnamefont{T.~M.} \bibnamefont{Truskett}},
  \bibinfo{author}{\bibfnamefont{S.}~\bibnamefont{Torquato}}, \bibnamefont{and}
  \bibinfo{author}{\bibfnamefont{P.~G.} \bibnamefont{Debenedetti}},
  \bibinfo{journal}{Phys. \ Rev. \ E} \textbf{\bibinfo{volume}{62}},
  \bibinfo{pages}{993} (\bibinfo{year}{2000}).

\bibitem[{\citenamefont{Kjellander and Sarman}(1991)}]{kjellander}
\bibinfo{author}{\bibfnamefont{R.}~\bibnamefont{Kjellander}} \bibnamefont{and}
  \bibinfo{author}{\bibfnamefont{S.}~\bibnamefont{Sarman}},
  \bibinfo{journal}{J. \ Chem. \ Soc. \ Faradau \ Trans.}
  \textbf{\bibinfo{volume}{87}}, \bibinfo{pages}{1869} (\bibinfo{year}{1991}).

\bibitem[{\citenamefont{Alcoutlabi and McKenna}(2005)}]{mckenna}
\bibinfo{author}{\bibfnamefont{M.}~\bibnamefont{Alcoutlabi}} \bibnamefont{and}
  \bibinfo{author}{\bibfnamefont{G.~B.} \bibnamefont{McKenna}},
  \bibinfo{journal}{J. \ Phys.: \ Condens. \ Matter}
  \textbf{\bibinfo{volume}{17}}, \bibinfo{pages}{R461} (\bibinfo{year}{2005}).

\bibitem[{\citenamefont{Sollich}(2002)}]{sollich}
\bibinfo{author}{\bibfnamefont{P.}~\bibnamefont{Sollich}}, \bibinfo{journal}{J.
  \ Phys.: \ Condens. \ Matter} \textbf{\bibinfo{volume}{14}},
  \bibinfo{pages}{R79} (\bibinfo{year}{2002}).

\end{thebibliography}

\end{document}